\documentclass[aps,prb,twocolumn,showpacs,superscriptaddress]{revtex4}
\usepackage{graphicx}
\usepackage{bm}
\newcommand{\beq}{\begin{equation}}
\newcommand{\eeq}{\end{equation}}

\begin{document}
\title{\boldmath  Two Scenarios of the Quantum Critical Point}
\author{V.~A.~Khodel}
\affiliation{Russian Research Center Kurchatov Institute, Moscow,
123182 Russia } \affiliation{McDonnell Center
for the Space Science and Department of Physics, \\
Washington University, St.Louis, MO 63130, USA  }

\vskip 0.8 cm

\date{\today}

\begin{abstract}
Two different scenarios of the quantum critical point (QCP), a zero-temperature instability of the Landau state, related to  the divergence of the effective mass, are investigated.
Flaws of the standard scenario of the QCP, where this divergence is attributed  to  the occurrence of  some second--order phase transition, are demonstrated. Salient features of a different
{\it topological} scenario of  the  QCP, associated with  the emergence of bifurcation points in  equation $\epsilon(p)=\mu$  that ordinarily determines the Fermi momentum, are analyzed. The topological  scenario of the QCP  is applied to  three-dimensional (3D) Fermi liquids with
an attractive  current-current interaction.
\end{abstract}

\maketitle

A statement  that the Landau quasiparticle picture breaks down at
points of  second--order phase transitions  has become a truism.
The violation of this  picture is attributed to  vanishing of the
quasiparticle weight $z$ in the single-particle state since the
analysis  of a long wave-length instability in the $S=1$
particle-hole channel, performed  more than forty years ago by
Doniach and Engelsberg  \cite{doniach} and refined later by
Dyugaev. \cite{dyugaev}  In nonsuperfluid Fermi systems, the
$z$-factor is determined by the formula
$z=[1-\left(\partial\Sigma(p,\varepsilon)/\partial\varepsilon\right)_0]^{-1}$
where the subscript $0$ indicates that the respective derivative
of the mass operator $\Sigma$ is evaluated at the Fermi surface.
This factor enters a textbook formula
\beq
   {M\over M^*}=z\left[1+\left({\partial\Sigma(p,\varepsilon)\over\partial\epsilon^0_p}\right)_0\right] \
\label{meffz}
\eeq
for the ratio  $M^*/M$ of the effective mass $M^*$ to the mass $M$ of a free particle. As seen from this formula,
where $\epsilon^0_p=p^2/2M$,
 the effective mass diverges at a critical density $\rho_c$, where   $z$ vanishes  provided the sum
 $1+\left(\partial\Sigma(p,\varepsilon=0)/\partial\epsilon^0_p\right)_0$ has a {\it  positive and finite value} at this point. Nowadays,
when studying  critical fluctuations of arbitrary wave-lengths
$k<2p_F$ has become popular, this restriction  is often assumed to
be met without stipulations. E.g.\ a standard scenario of the quantum critical point (QCP) where $M^*$ diverges
is formulated as follows: in the vicinity of an impending
second--order phase transition,
 "quasiparticles get heavy and die``.\cite{coleman1,coleman2}

However, as seen from Eq.(\ref{meffz}), $M^*$ may diverge not only at the points of the second-order phase transitions, but also at a critical density $\rho_{\infty}$, where the sum $1+\left({\partial\Sigma(p,\varepsilon)\over\partial\epsilon^0_p}\right)_0$ changes its sign. Furthermore, we will demonstrate that except for the case of the ferromagnetic instability,\cite{doniach} $M^*$ cannot diverge at $\rho_c$ without violation of  stability conditions.

In what follows we   restrict ourselves to one-component  three-dimensional (3D) homogeneous Fermi liquids where the particle momentum is conserved, and
 the Landau equation, connecting the quasiparticles group velocity
  $\partial\epsilon/\partial{\bf p}$
to their momentum distribution $n(p)$
in terms of the interaction function $f$, has the form \cite{lan,lanl,trio}
\beq
{\partial\epsilon(p)\over\partial {\bf p}} =
  {{\bf p}\over M} +
     \int\! f({\bf p},{\bf p_1})\,
         {\partial n(p_1)\over\partial {\bf p_1}}\, {d^3p_1\over (2\pi)^3}  \  .
\label{lan}
\eeq
  Setting here $p=p_F$ and introducing the notation $v_F=\left(d\epsilon(p)/dp \right)_0=p_F/M^*$, one obtains
   \beq
    v_F={p_F\over M}\left(1-{p_FM\over 3\pi^2}f_1\right)  \ ,
    \label{gr}
    \eeq
    implying that
\beq
{M\over M^*}=1-{1\over 3} {p_FM\over \pi^2} f_1\   .
\label{meffl}
\eeq
Hereafter we employ   notations of  Fermi liquid (FL) theory  where $f_1$ is the first harmonic of  the  interaction function $f(\theta)=z^2\Gamma^{\omega}(p_F,p_F;\theta)$, with
  $\Gamma^{\omega}$ being
   the  $\omega$-limit of  the
scattering amplitude $\Gamma$ of two particles, whose
energies  and  incoming momenta ${\bf
p}_1,{\bf p}_2$ lie on the Fermi surface, with $\cos\theta={\bf p}_1\cdot{\bf p}_2/p^2_F$, while the 4-momentum
transfer $({\bf q},\omega)$ approaches zero, such that $q/\omega\to 0$.

 It is instructive to rewrite Eq.(\ref{meffl}) in terms of the  $k$-limit of the dimensionless scattering amplitude $\nu\Gamma^k=A+B
 {\bm \sigma}_1{\bm \sigma}_2$ where $\nu =z^2p_FM^*/\pi^2$ is the quasiparticle density of states.  Simple algebra then yields
 \beq
 {M\over M^*}=1-{1\over 3} A_1  \  .
 \label{meffm}
 \eeq
   This formula stems from  Eq.(\ref{meffl})  and  relation \cite{lan,lanl,trio}
 $ A_1=\Phi_1/ (1+{\Phi_1\over 3} )$,
  where $\Phi$ is the spin-independent part of the  product
  $\nu \Gamma^{\omega}$. Thus at the density $\rho_{\infty}$ where the effective mass diverges one has
  \beq
  A_1(\rho_{\infty})=3 \  ,\quad  \Phi_1(\rho_{\infty})=\infty  \   .
 \eeq

\begin{figure}[t]
\includegraphics[width=0.7\linewidth,height=0.22\linewidth]{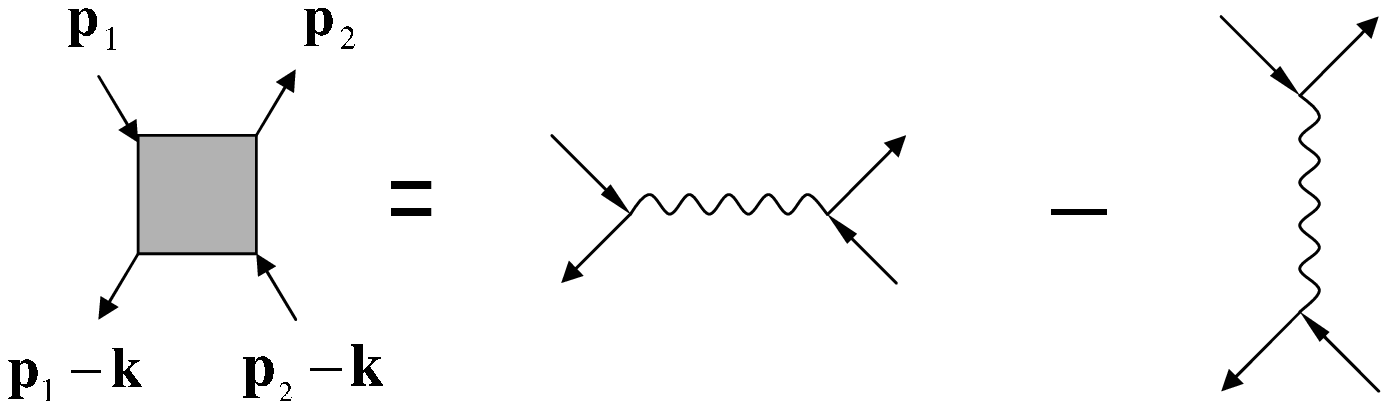}
\caption {Singular contributions to the scattering amplitude in
the vicinity of the second--order phase transition.}
\label{fig:figure}
\end{figure}

 In the following we focus on  critical density fluctuations with $k_c\neq 0$, addressed
in Ref.~\onlinecite{chubukov}. First we notice that
  there is  a strong dependence of  the amplitude  $\Gamma_{\alpha\beta,\gamma\delta}({\bf p}_1,{\bf p}_2,{\bf k},\omega=0)$
  on the momentum transfer $k$ close to the critical momentum $k_c$, specifying the spectrum of  density fluctuations, that stems
   from
 the asymmetry of  $\Gamma$ with respect to the interchange of
 momenta and spins of colliding particles. \cite{dyugaev}
 In this case, upon neglecting regular  components  one finds\cite{dyugaev},
(see Fig.~\ref{fig:figure}),
   \beq
  A({\bf p}_1,{\bf p}_2,{\bf k},\omega=0;\rho\to \rho_c)=-D({\bf k})+{1\over 2} D({\bf p}_1-{\bf p}_2+{\bf k})  \  ,
  \label{dyug}
  \eeq
  with
  \beq
 D(k\to k_c,\omega=0)={g\over \xi^{-2}(\rho) +(k-k_c)^2}  \  ,
 \label{crit}
 \eeq
   the correlation length $\xi(\rho)$ diverging at $\rho=\rho_c$.

  Within the  quasiboson approximation \cite{chubukov}, the derivative $\left(\partial\Sigma(p,\varepsilon)/\partial\varepsilon\right)_0$ diverges   at $\rho\to \rho_c$ as $ \xi(\rho)$, while the derivative  $\left(\partial\Sigma(p,\varepsilon)/\partial\epsilon^0_p\right)_0$ remains finite at any  $k_c$. If these results were correct, then the densities $\rho_{\infty}$ and $\rho_c$ would coincide,  in agreement with the standard scenario of the QCP.   However,  calculations of
 harmonics $A_k(\rho)$ of the  amplitude $A(p_F,p_F,\cos\theta)$ from  Eqs.(\ref{dyug}) and (\ref{crit}) yield
 \beq
 A_0(\rho{\to}\rho_c)=g{\pi\over 2}{k_c\xi(\rho)\over p^2_F}, \;
 A_1(\rho{\to}\rho_c)=g{3\pi\over 2}
 {k_c\xi(\rho)\over p^2_F}\cos\theta_0.
 \label{harm}
 \eeq
We see that the sign of $A_1(\rho\to \rho_c)$,  coinciding with that of $\cos\theta_0=1-k^2_c/2p^2_F$, turns out to be {\it negative} at $k_c>p_F\sqrt{2}$. According to  Eq.(\ref{meffm}) this implies that at the  point of the second--order phase transition, the ratio $M^*(\rho_c)/M<1$. Thus we infer that at  $k_c>p_F\sqrt{2}$, the densities $\rho_c$ and $\rho_{\infty}$ {\it cannot coincide}.
  In its turn,  this implies that vanishing of the $z$-factor at $\rho_c$ is compensated by the divergence of the derivative $(\partial\Sigma(p,\varepsilon)/\partial\epsilon^0_p)_0$   at this point,  otherwise Eq.(\ref{meffz}) fails.

  To verify this assertion let us write down a fundamental FL relation \cite{trio} between the $k$- and $\omega$-limits of the vertex ${\cal T}$ that has the symbolic form ${\cal T}^k={\cal T}^{\omega}+\left(\Gamma^k ((G^2)^k-(G^2)^{\omega}){\cal T}^{\omega}\right)$ where external brackets mean integration and summation over all intermediate momemtum and spin variables. In dealing with the bare vertex ${\cal T}^0={\bf p}$  the extended form of this relation is
 $$
 -{\partial G^{-1}(p)\over \partial {\bf p}}={\partial G^{-1}(p)\over \partial \varepsilon} {{\bf  p}\over M}+$$
 \beq
 Sp_{\sigma}\int \Gamma^k(p,q)\left(\left(G^2(q)\right)^k-\left(G^2(q)\right)^{\omega}\right){\partial G^{-1}(q)\over \partial q_0} {{\bf  q}\over M}
 {d^4q\over (2\pi)^4i}  \  ,
 \label{kom}
 \eeq
  In writing this equation
  the Pitaevskii identities\cite{trio}
  \beq
  {\cal T}^k({\bf p})=-{\partial G^{-1}(p,\varepsilon)\over \partial {\bf p}}\  ,\quad {\cal T}^{\omega}({\bf p})={\partial G^{-1}(p,\varepsilon)\over \partial \varepsilon}{\bf p}  \
  \label{pit}
  \eeq
  are employed. Upon inserting the FL formula
    \beq
  \left(G^2(q)\right)^k=  \left(G^2(q)\right)^{\omega}-2\pi^3i{\nu\over p^2_F}\,\delta(\varepsilon)\delta(p-p_F)
  \eeq
 into Eq.   (\ref{kom}) and the standard replacement of  the spin-independent part of $\nu\Gamma^k$  by $A$,  after some algebra we are led to equation
  \beq
 1+ \left({\partial \Sigma(p,\varepsilon)\over \partial \epsilon^0_p}\right)_0=\left(1-\left({\partial\Sigma(p,\varepsilon)\over \partial\varepsilon}\right)_0\right)\left(1-{1\over 3} A_1\right) \  .
 \label{rel1}
 \eeq
 Remembering that $-{\partial G^{-1}(p,\varepsilon)/ \partial {\bf p}}=z^{-1}d\epsilon(p)/dp$ one arrives \cite{trio} at  Eq.(\ref{meffm}). On the other hand,
 as seen from Eq.(\ref{kom}), at $k_c>p_F\sqrt{2}$ where  $A_1<0$,  the derivative $\left(\partial \Sigma(p,\varepsilon)/ \partial \epsilon^0_p\right)_0$ does diverge at the same density, as the derivative $\left(\partial \Sigma(p,\epsilon)/ \partial \varepsilon\right)_0$, in contrast to the result.\cite{chubukov}
 To correct the defect of the quasiboson approximation\cite{chubukov}, the spin-independent part  of the scattering amplitude $\Gamma$ entering  the formulas\cite{chubukov}  for the derivatives of the mass operator $\Sigma$  should be replaced by that of the amplitude $\Gamma^{\omega}$ (for details,
see Ref.~\onlinecite{dyugaev}).

 We will immediately see that at finite $k_c<p_F\sqrt{2}$, vanishing of the $z$-factor is {\it incompatible} with the divergence of  $M^*$  as well. Indeed, as seen from Eq.(\ref{harm}),  the harmonics $A_0(\rho_c)$ and $A_1(\rho_c)$ are related to each other by equation $A_0(\rho_c)=A_1( \rho_c)/(3\cos\theta_0)$. If  $M^*(\rho_c)$ were infinite, then according to Eq.(\ref{meffm}), $A_1(\rho_c)$  would equal  3, and $A_0(\rho_c)=1/\cos\theta_0$.  However, the quantity $A_0=\Phi_0/(1+\Phi_0)$ cannot be in excess of 1, otherwise  the Pomeranchuck stability condition \cite{lanl,trio} $\Phi_0>-1$ is violated, and the compressibility turns out to be negative. Thus  the QCP cannot be reached without the violation of the stability condition. If so, approaching the QCP, the system  undergoes a first--order phase transition, as in the case of 3D liquid $^3$He.

     The finiteness of  $M^*$ at the points of vanishing of the $z$-factor  requires an alternative explanation,
(see e.g.\ Ref.~\onlinecite{zk}), of  the logarithmic enhancement of the specific heat $C(T)$, observed in many heavy fermion metals\cite{gegenwart}
    and attributed to  contributions of critical fluctuations.
\begin{figure}[t]
\includegraphics[width=0.7\linewidth,height=0.54\linewidth]{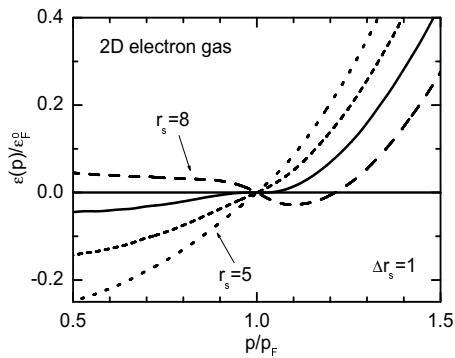}
\caption
{Single-particle spectrum $\epsilon(p)$ of a homogeneous 2D
 electron gas in units of $\epsilon_F^0=p^2_F/2M$,
evaluated \cite{bz} at $T=0$ for different values of the dimensionless
 parameter $r_s=\sqrt{2}Me^2/p_F$.}
\label{fig:fc_2deg}
\end{figure}

 Let us now  turn to the analysis of another opportunity for  the occurrence of the QCP, addressed first in microscopic calculations of the single-particle spectrum of 3D electron gas.\cite{ks1,zjetp} It is  associated with the change of the sign of the sum  $1+ \left({\partial \Sigma(p,\varepsilon)\over \partial \epsilon^0_p}\right)_0$ at  $\rho_{\infty}\neq \rho_c$. In this case, the $z$-factor keeps its finite value, and hence, the quasiparticle picture holds on both the sides of the QCP.
 In standard Landau theory,  equation
 \beq
 \epsilon(p,T=0)=\mu
 \label{bif}
 \eeq
 with $\mu$, being  the chemical potential,
  has the single root, determining the Fermi momentum $p_F$. Suppose, at a critical coupling constant $g_T$, a bifurcation in Eq.(\ref{bif}) emerges, then beyond the critical point, at $g>g_T$, this equation acquires, at least, two new roots that triggers a {\it topological} phase transition.\cite{volrev}
   In many-body theory,  equation, determining  critical  points of the topological phase transitions, has the form
   \beq
   \epsilon^0_p+\Sigma(p,\varepsilon=0)=\mu  \  .
   \label{topm}
   \eeq
 Significantly, terms,  proportional to  $\epsilon\ln\epsilon$, existing  in  the mass operator $\Sigma$ of  marginal Fermi liquids, do not enter this equation.

  The bifurcation $p_b$ in Eqs.(\ref{bif}) and (\ref{topm}) can emerge at any point of momentum space.  If $p_b$ coincides with the Fermi momentum $p_F$, then
  at the critical density the sum $1+ \left({\partial \Sigma(p,\varepsilon)\over \partial \epsilon^0_p}\right)_0$  vanishes, and one arrives at
    the {\it topological} quantum critical point. In connection with this scenario,
     it is instructive to trace the evolution of the group velocity  $v_F=(d\epsilon/dp)_0$ versus  the first harmonic $f_1$.
    As follows from   Eq.(\ref{gr}), $v_F$ keeps its positive sign as long as $F^0_1=p_FMf_1/3\pi^2<3$,  and the Landau state with the quasiparticle momentum distribution $n_F(p)=\theta(p-p_F)$ remains stable.
   However, at  $F^0_1> 3$, the sign of $v_F$ changes, and the Landau state is necessarily rearranged.
    This conclusion is in agreement with results of microscopic calculations of the single-particle spectrum $\epsilon(p,T=0)$ of 2D electron gas, \cite{bz} shown in  Fig.~\ref{fig:fc_2deg}. As seen, the sign of $v_F$ holds until the dimensionless parameter $r_s$ attains a critical value $r_{sc}\simeq 7.0$. At greater $r_s$, the derivative $\left(d\epsilon(p)/dp\right)_0$, evaluated  with the momentum distibution $n_F(p)$, becomes negative, and the Landau state loses its stability, since the curve $\epsilon(p)$ crosses the Fermi level more than one time.

    At $T=0$, two types of the topological transitions are known. \cite{volrev}  One of them, giving rise to
  the multi-connected Fermi surface,   was uncovered \cite{zb} and studied later \cite{shagp,vosk,schofield,zk}
   in calculations of the single-particle spectrum $\epsilon(p)$ on the base of Eq. (\ref{lan}), with  the interaction function $f(k)$, having   no singularities at $k\to 0$.  In this case, beyond the
 QCP, Eq.(\ref{bif})
  has three roots $p_1<p_2<p_3$, i.e. the curve $\epsilon(p)$ crosses the Fermi level
three times, and    occupation numbers  are:
$n(p)=1$ at $p<p_1$,  $n(p)=0$ at $p_1<p<p_2$, while at $p_2<p<p_3$, once again $n(p)=1$, and at $p>p_3$, $n(p)=0$.
  As the  coupling constant $g$ increases, the number of the roots of Eq.(\ref{bif})
 rapidly grows,  however, their number  remains countable at any $g>g_T$.

The situation changes  in     Fermi liquids with a singular  attractive  long-range current-current term
  \beq
 \Gamma^0({\bf p}_1,{\bf p}_2,{\bf k},\omega=0)=-g {{\bf p}_1{\bf p}_2-({\bf p}_1{\bf k})({\bf p}_2{\bf k})/k^2
 \over k^2}  \  ,
  \label{cur}
 \eeq
  since in these systems,  e.g.  in dense quark-gluon plasma, solutions with the multi-connected Fermi surface are unstable.  Indeed, the group velocity $d\epsilon(p)/dp$ evaluated with  $n_F(p)=\theta(p-p_F)$  from  Eq. (\ref{lan}) has the form $d\epsilon(p)/dp=p_F/M-g \ln(2p_F/|p_F-p|)$,  implying that Eq.(\ref{bif}) has three different roots $p_1,p_2,p_3$, corresponding to   the Fermi surface, having   three sheets at any $g>0$. \cite{baym} However, at the next iteration step,   the new  Fermi surface
has already five sheets, the  Fermi surface has
seven sheets  and so on. \cite{baym} With increasing the
number of iterations, the distance between neighbour sheets  rapidly
shrinks. In this situation, a minute elevation of temperature renders the
momentum distribution $n(p,T)$ a smooth  $T$-{\it independent} function $n_*(p)$, different from 0 and 1 in a domain ${\cal C}$ between the sheets.
In this case, the ground--state stability condition,
\beq
\delta E=\int \left(\epsilon(p)-\mu\right)\delta n(p){d^3p\over (2\pi)^3}>0  \  ,
\eeq
 requiring the nonnegativity of the variation $\delta E$ of the ground state energy $E$ at any admissable variation of $n_*(p)$, is met provided
 \beq
 \epsilon(p)=\mu \ ,\quad p\in {\cal C}  \  .
 \label{fct}
 \eeq
As a result,  we  arrive at another type of the
topological  transitions, the so-called fermion condensation \cite{ks,vol,noz,physrep,yak,volrev,shagrev}, where the
roots of Eq.(\ref{bif}) form an {\it uncountable} set, called the fermion condensate (FC).
Since the quasiparticle energy $\epsilon(p)$ is nothing but the derivative of the ground state energy $E$ with respect to the quasiparticle momentum distribution $n(p)$, Eq.(\ref{fct})  can be rewritten as
 variational condition \cite{ks}
\beq
{\delta E\over \delta n(p)}=\mu    , \quad p \in {\cal C}  \  .
\label{var}
\eeq

The FC Green function has the form
    \beq
    G(p,\varepsilon)={1-n_*(p)\over \varepsilon+i\delta}+ {n_*(p)\over \varepsilon-i\delta}\  ,\quad  p\in {\cal C}  \  .
    \eeq
    As seen, only the imaginary part of  the FC Green function differs from that of  the ordinary FL Green function. This difference exhibits itself in a {\it topological  charge}, given by the integral\cite{vol,volrev}
   \beq
   N=\int\limits_{\gamma}  G(p,\varepsilon)\, \partial_l G(p,\varepsilon) {dl\over 2\pi i}  \  ,
 \eeq
 where integration is performed over a contour in complex energy plane, embracing the Fermi surface. For conventional Fermi liquids and systems with the multi-connected Fermi surface, the topological charge $N$ is integer, while for the states with a FC, its value is {\it half-integer}.\cite{vol,volrev}

For illustration of the phenomenon of fermion condensation, let us address   dense quark-gluon plasma,    on the Lifshitz phase diagram of which, as we have seen,  there is no room for the conventional FL phase.\cite{khv}   Upon inserting into Eq.(\ref{lan}) only leading divergent terms  in the interaction function $f$, constructed from Eq.(\ref{cur}), one finds
   \beq
   0= 1-\lambda\int \ln {1\over |x-x_1|} {\partial n_*(x_1)\over \partial x_1} dx_1\   , \quad x,x_1\in {\cal C}  \  ,
   \label{eqfc}
   \eeq
  where  dimensionless variables $x=(p_F-p)/2p_F$ and  $\lambda$ are introduced. A numerical solution of this equation will be found elsewhere. Here  we  simplify Eq.(\ref{eqfc}), replacing  the kernel $\ln (1/|x-x_1|)$ by $\ln (1/x) $ provided $x>x_1$, and by $\ln (1/x_1)$, otherwise, to obtain
  \beq
  0=1+\lambda n_*(x)\ln x +\lambda\int\limits_x^{x_m}\ln x_1{\partial n_*(x_1)\over \partial x_1} dx_1 \,   \quad x,x_1\in {\cal C} .
  \label{eqfc1}
  \eeq
  An approximate solution of this equation is $n_*(x)=x/ x_m$ where $x_m=e^{-{1\over \lambda}}$ is determined from condition $n(x_m)=1$. We see that the range of the interval $[0,x_m]$ of fermion condensation, adjacent to the Fermi surface, is  exponentially small.

 Nontrivial smooth solutions $n_*(p)$ of Eq.(\ref{var}) exist even in weakly correlated Fermi systems. However,  in these systems, the Pauli restriction
  $n_*(p)<1$ is violated, rendering such solutions  meaningless.  Even at the QCP, where the nonsingular interaction function $f$
  is already sufficiently strong,  no consistent FC solutions $n_*(p)$  exist, satisfying the requirement $n(p)<1$ wherever. These solutions emerge at
  a critical  constant $g_{FC}$, and at $g>g_{FC}$, they win the contest with any other solutions.
  Thus on the Lifshitz phase diagram of systems with   nonsingular repulsive interaction functions $f$, the standard FL phase occupies the interval $g<g_T$, the phase with the multi-connected Fermi surface, the interval $g_T<g<g_{FC}$, while  the phase with the FC exists at $g>g_{FC}$.

  In dealing with the full $(T-g)$ phase diagram of such systems we notice that the temperature evolution of the quasiparticle momentum distribution, associated at $T=0$ with  the  multi-connected  Fermi surface,  depends on   the departure of the difference $|\epsilon(p)-\mu|$ from 0 in the domain ${\cal C}$.  Its  maximimu value $\epsilon_m$ determines a new energy scale $\epsilon_m\simeq d^2/M^*(0)$, where $d$ is the average distance between the sheets of the Fermi surface that  rapidly falls  with the increase of the sheets number.\cite{zb,shagp} If temperature $T$ attains values, comparable with $\epsilon_m$, then, as seen from the Landau formula $n(p)=[1+\exp\left(\epsilon(p)-\mu)/T\right)]^{-1}$, the distribution $n(p)$  becomes a smooth function of $p$. Employing the  FC notation  $n_*(p)$
 for this function, one finds that at $T\geq \epsilon_m$  the spectrum
 \beq
 \epsilon(p,T)=T \ln {1-n_*(p)\over n_*(p)} \  , \quad p\in {\cal C}
 \eeq
  does coincide with  the FC spectrum \cite{noz}.  We infer that at $T\simeq \epsilon_m$, a crossover from a state with the multi-connected Fermi surface to a state with the FC occurs.  As a result, FL thermodynamics of the systems with the multi-connected Fermi surface completely alters at $T\simeq \epsilon_m$, since properties of systems with the FC resemble those of a gas of localized spins.\cite{zk1}   Such a transition was recently observed
  in the heavy--fermion metal YbIr$_2$Si$_2$, transition temperature  being merely 1 K.  \cite{steglich1}

   Let us now turn to systems
   of fermions, interacting with a "foreign`` bosonic mode, e.g. phonons or photons. In the Fr\"olich model, \cite{trio,mig} aimed for the elucidation of  electron and phonon spectra in solids,   electrons share momentum with the lattice  due to the electron-phonon interaction.   The non-conservation of the electron momentum results in the {\it violation} of the second of the  relations (\ref{pit}), and Eq.(\ref{meffl}) acquires the form
   \beq
   {M\over M^*}=z {\left({\bf n}{\cal T}^{\omega}({\bf p})\right)_0\over p_F}\left(1-{1\over 3} A_1\right) \  ,
   \label{meffr}
   \eeq
   where ${\bf n}={\bf p}/p$.   The departure of
    the ratio $M/M^*$ from the Landau value (\ref{meffl}), is well pronounced in the limit $c_s<<v_F$.  For illustration, let us consider the weak coupling limit of   the Fr\"olich model, where the first harmonic $A_1$, evaluated from the phonon propagator $D(|{\bf p}_1-{\bf p}_2|,\omega=0)$ equals 0 due to its isotropy. If Eq.(\ref{meffl}) were correct, then $M^*/M$ would equal 1. However, this is not the case:    $M^*/M=1+g^2p_FM/2\pi^2$ where $g$ is the electron-phonon coupling constant.\cite{mig}  Evidently, if the ratio $v_F/c_s$ drops, then the departure from Eq. (\ref{meffl}) falls due to weakening of the contribution of the pole of the boson propagator. Such a situation occurs just in the vicinity of the QCP, since at this point $v_F=p_F/M^*=0$.

So far information on the QCP properties of  Fermi liquids  is extracted from measurements, carried out in 2D electron gas,  2D liquid $^3$He and heavy--fermion metals.  Here we restrict ourselves to several remarks, reserving a more detailed analysis  for a separate paper.  Accurate measurements  of the effective mass $M^*$ in dilute
2D electron gas are made on (100)- and (111)-silicon MOSFET's. \cite{shashkin-PRB,pudalov-PRL,shashkin1,shashkincond} In principle, the divergence of $M^*$, observed in these experiments, can be  associated with critical spin-density fluctuations. However,   experimental data rules out a significant enhancement of the  Stoner factor. In many theories,
(see e.g.\ Ref.~\onlinecite{fink}), the enhancement of $M^*$ is related to disorder effects. However, the effective masses, specifying the electron spectra of (100)- and (111)-silicon MOSFET's, where disorder is different, almost coincide with each other provided  dimensionless parameters $r_s$ of the 2D Coulomb problem, are the same.\cite{shashkincond} On the other hand, this coincidence that agrees with results of microscopic calculations \cite{bz}  is straightforwardly elucidated within
  the topological scenario of the QCP.
   There are  reports on the divergence of the effective mass in 2D liquid $^3$He,
  (see e.g.\ Refs.~\onlinecite{godfrin1,godfrin2,saunders-PRL,saunderssci}).  Furthermore,
following \cite{godfrin2}, authors
of Ref.~\onlinecite{saunderssci} reported that at the density
$\rho>9.00nm^{-2}$, the low-temperature limit of the product $T\chi(T)$
quickly increases with increasing $\rho$. In addition,  the ratio  of the specific heat $C(T)$ to $T$ does not obey FL theory in this density region, since it increases with lowering $T$.  These facts can be interpreted as evidence for the presence of the FC.\cite{yak} Unfortunately, so far the accuracy of extremely difficult measurements of the ratio $C(T)/T$ at $T\leq 1K$, is insufficient to properly evaluate a low-temperature  part of the entropy $S$ and compare it with the respective FC entropy, extracted from data on $\chi(T)$.
   The divergence of the ratio $C(T)/T$, associated with the QCP, is observed in several heavy--fermion compounds.\cite{gegenwart,steglich2,steglich3}
   Authors of the experimental article \cite{steglich3} claim that  data on the  Sommerfeld-Wilson ratio $R_{SW}=\chi(T)/C(T)$  in a doped compound
    YbRh$_2$(Si$_{0.95}$Ge$_{0.05}$)$_2$ point to an enhancement of the Stoner factor that has to be infinite  at the point of the ferromagnetic phase transition.  However,  evaluation of the Stoner factor from experimental data in heavy--fermion metals encounters difficulties, discussed
in Ref.~\onlinecite{zk1}. Furthermore,  with a correct normalization experimental data \cite{steglich3} are explained without any enhancement of the Stoner factor \cite{zk1} that rules out the relevance of the ferromagnetic phase transition to the QCP in this metal. A different
     H--T phase diagram is constructed for the heavy--fermion metal  YbAgGe   in
Refs.\onlinecite{budko}. On this diagram, the FL phase  is separated from  a phase with  magnetic  ordering by a  significant domain  of NFL  behavior. Such a separation is easily explained within the topological scenario of the QCP, discussed in this article.

In conclusion, in this article, two different scenarios of the quantum critical point (QCP), a low-temperature instability of the Landau state, related to
the divergence of the density of states $N(0)\sim M^*$, are analyzed. We  discuss shortcomings of  the conventional scenario of the QCP, where the divergence of  the effective mass $M^*$ is attributed to vanishing of the quasiparticle weight in the single-particle state. In a different, topological scenario, associated with the change of the topology of the Fermi surface at the QCP, the quasiparticle picture holds on both the sides  of the QCP.   This scenario is in agreement with microscopic calculations of the QCP in 2D electron gas and  does not contradict relevant experimental data on  2D liquid $^3$He and heavy--fermion metals.

I thank A.~S. Alexandrov, G. Baym, S. L. Bud'ko, V. T. Dolgopolov,
J. Quintanilla, M. R. Norman, S. S. Pankratov, A. Schmitt, G. E. Volovik  and M. V. Zverev for fruitful  discussions.
This research was supported by the McDonnell Center for the Space
Sciences, by Grant No.~NS-8756.2006.2 from the Russian Ministry of
Education and Science and by Grants Nos.~06-02-17171-a and
07-0200553-a from the Russian Foundation for Basic Research.

\end{document}